\documentclass[aps,showpacs,amssymb,amsmath,prl,twocolumn,superscriptaddress]{revtex4}

\usepackage{bm}
\usepackage[latin1] {inputenc}
\usepackage{graphicx}
\begin{document}

\title{Dynamic stability of spindles controlled by molecular motor kinetics}

\author{Otger Camp\`as}\affiliation{Institut Curie, UMR 168, 26 rue d'Ulm, F-75248 Paris Cedex
05, France}\affiliation{Departament d'ECM, Universitat de
Barcelona, Avinguda Diagonal 647, E-08028 Barcelona, Spain}

\author{Jaume Casademunt}\affiliation{Departament d'ECM, Universitat de
  Barcelona, Avinguda Diagonal 647, E-08028 Barcelona, Spain}

\author{Ignacio Pagonabarraga}\affiliation{Departament de F\'isica Fonamental, Universitat de
  Barcelona, Avinguda Diagonal 647, E-08028 Barcelona, Spain}


\begin{abstract}
We analyze the role of the force-dependent kinetics of motor
proteins in the stability of antiparallel arrays of polar
filaments, such as those in the mitotic spindle. We determine the possible
stable structures and show that there exists an
instability associated to the collective behavior of motors that
leads to the collapse of the spindle. Our analysis provides a general
framework to understand several experimental observations in
eukaryotic cell division.
\end{abstract}

\maketitle

Living cells display several structures that arise from the self-organization
of polar filaments and motor proteins~\cite{Alberts}. Several \emph{in vitro} studies
have shown that mixtures of kinesin motors and 
microtubules (MTs) can spontaneously develop complex spatio-temporal
patterns~\cite{Nedelec01}. These self-organization processes are
essential for eukaryotic cell division~\cite{Wittmann}. During mitosis, motor
proteins organize MTs in a bipolar structure, 
the mitotic spindle, which serves as a scaffold to transmit the
necessary forces for chromosome segregation~\cite{Salmon_Mitch}.
The spindle consists of two MT asters that overlap in the central region. The
MTs, with their minus-ends located at the aster poles, 
are crosslinked by many different motor proteins~\cite{Sharp,Wittmann}.
One particular type of motors, the plus-ended bipolar kinesins (e.g. Eg5  or
Klp61F), has been shown to be essential for the spindle stability. A decrease
in their concentration below a certain threshold causes the spindle
collapse~\cite{Kapoor,Mitchison_eg5in}, and their total inhibition prevents
bipolar spindle formation~\cite{Vale}. In addition, 
Eg5 motors have been shown to drive the MT poleward
flux~\cite{Mitchison_eg5in} and homolog motors to induce the
formation of (interpolar) MT bundles~\cite{Sharp02}.

\begin{figure}[!h]
{\centering\resizebox*{8.5cm}{!}{\includegraphics{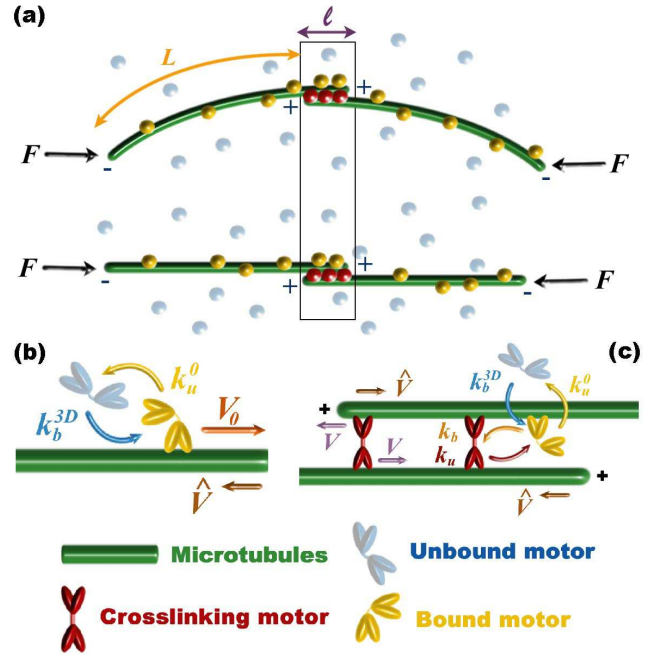}}}
\caption{\label{sketch} (a) Antiparallel array of
  MTs under the action of a longitudinal force $F$. Buckled and
  non-buckled arrays are shown. The minus and plus ends of MTs are depicted as
  $-$ and $+$ respectively. Motors are represented by dots. In the non-overlapping region of length $L$ there are no
  antiparallel filaments and the motors are not subject to any force. The motors in
  the overlapping region of length $\ell$ sustain the structure by crosslinking and sliding
  antiparallel filaments. (b-c) Possible kinetic events of a motor in the
  non-overlapping region (b) and in the overlapping region (c). The velocities
  $V$ and $\hat{V}$ are the crosslinking motor velocity and the MT poleward
  velocity respectively.}
\end{figure}

Bipolar motors are composed of two connected units, each one composed of two
motor domains. Both units can move simultaneously and independently on
MTs~\cite{Schmidt}. These 
motors are able to crosslink MTs~\cite{Sharp02} and slide them with respect to
each other when they are in an antiparallel 
configuration~\cite{Schmidt}, like in the central region of the spindle
(Fig.~\ref{sketch}a,c). As a result, these motors produce an outward
force along the spindle axis and generate a MT flux toward the
poles~\cite{Mitchison_eg5in}. Typical
forces involved in mitosis lay in the nanoNewton range
~\cite{Nicklas}. Since individual motors cannot exert forces larger than a
few picoNewtons, their collective action is required to ensure the
stability of the mitotic spindle.
At metaphase, this dynamic structure reaches a steady
state with MTs of nearly constant length undergoing permanent
treadmilling~\cite{Mitchison_bipolar_flux,Mitchison_eg5in}, polymerizing
at the $+$ end and depolymerizing at the $-$ end.

The theoretical study of motors and MT mixtures has been recently addressed
 using continuum coarse grained descriptions~\cite{Frank,Jacques}, which have
 elucidated their basic self-organizing principles. 
 However, the coupling between force-dependent motor kinetics and local forces in self-organized
structures has not been addressed. In this Letter, we study the dynamic stability of
antiparallel arrays of MTs under the action of longitudinal forces,
in the presence of molecular motors able to collectively hold the structure by
stochastically crosslinking the filaments. We analyze the effects of the motor
 kinetics on the stability of the structure, and show that several
 phenomena observed in eukaryotic cell division can be understood on the same
 physical basis. 

In order to comprehend the basic physical mechanisms controlling the stability
of a spindle, we concentrate on a simplified geometry. We consider a pair of
antiparallel MTs (or an antiparallel MT
bundle) of fixed length, under the action of an inward force $F$
(Fig.~\ref{sketch}a). In the steady state,
there is a region of length $\ell$
where the antiparallel filament array overlaps (overlapping region). The
motors in this region can crosslink antiparallel MTs and slide them in opposite directions, generating
an outward force that balances the applied force $F$. We assume the antiparallel
MT sliding to be the only mechanism generating the poleward MT flux, as suggested
experimentally~\cite{Mitchison_eg5in}. Out of this overlapping region there
are two regions of length $L$ (non-overlapping regions; see
Fig.~\ref{sketch}a) where motors cannot apply
forces to sustain the spindle. Since we concentrate on the spindle
stability and do not address the mechanisms that determine $L$ and $\ell$, we
take them as given parameters. 

The motors in the non-overlapping region can be either bound to a
MT or freely diffusing in the bulk (Fig.~\ref{sketch}b). We assume a constant
bulk motor density, $\rho_{3D}$~\cite{assumption}, and consider the motors in
the bulk to attach onto MTs at a rate $k_b^{3D}$. Once bound to a MT, a
motor moves convectively with a mean velocity
$V_0$ toward the plus-end of the MT and detaches at a rate
$k_u^0$. At mean field level, the dynamics of the bound motor density,
$\rho(s,t)$, can then be expressed as~\cite{Lipowsky,Parmeggiani}
\begin{equation}
\partial_t \rho(s,t) + \partial_s J(s,t)  = - k_u^0 \rho(s,t) + k_b^{3D} \rho_{3D} \;,
\label{motordyn}
\end{equation}
where $s$ is the position along the MT as measured from the MT minus ends and $J(s,t)$ is the flux
of bound motors. For simplicity, we assume the bound motors to be in
a low density phase and write $J(s,t) = \rho(s,t) ( V_0-\hat{V})$ in the laboratory reference
frame, with $\hat{V}$ being the velocity of MTs toward the poles (Fig.~\ref{sketch}b,c).

In the central overlapping region, bipolar motors can be either in a
crosslinking state or in a bound state. In the former state both motor units 
are attached to a pair of antiparallel MTs, sliding them in opposite
directions and supporting a fraction of the total force $F$. As a result, the crosslinking motors move with a
force-dependent velocity $V$ (Fig.~\ref{sketch}c). Based on experimental
observations~\cite{Block}, we write a linear force-velocity relation, $V=V_0
\left( 1-f_m/f_s\right)$, where 
$f_m$ is the load applied on the motor and $f_s$ its stall force. We consider
a number $n_c$ of independent crosslinking motors to equally share the total applied
force, so that $f_m=F/n_c$. As the
poleward MT movement is driven only by these motors, we identify
$\hat{V}=V$ (Fig.~\ref{sketch}c). Each unit of a motor in the crosslinking state can detach at a
force-dependent rate $k_u(f_m) = k_u^0 \exp( f_m b/K_B T)$ (Kramers
theory~\cite{VanKampen}), where $b$ is a length
in the nanometer scale characterizing the activated process and $K_BT$ the thermal
energy. Such exponential sensitivity to applied load has indeed been observed
experimentally~\cite{Veigel}. After the detachment of one motor unit, the
bipolar motor is only 
bound to one MT and unable to apply force. Such motor can either detach the
bound motor unit left at a rate $k_u^0$ and diffuse into the bulk, or re-attach the unbound motor
unit at a rate $k_b$ and become a crosslinking motor again. The motors in
the bulk can also attach directly to the MTs in the overlapping region at a
rate $k_b^{3D}$.

The relevant variables being the  number of motors sustaining the spindle, we
neglect their spatial distribution in the overlapping region. Accordingly, the
equations for the average number of crosslinking and bound motors, $n_c$ 
and $n_b$ respectively, read
\begin{eqnarray}
\label{eq1}
\frac{dn_c}{dt} &=& k_b n_b - k_u(n_c) n_c \;,\\
\frac{dn_b}{dt} &=& 2 J(L,t) + k_b^{3D}\rho_{3D} \ell + k_u(n_c) n_c -
\left( k_u^0 + k_b \right) n_b  \;, \nonumber
\end{eqnarray}
where $J(L,t)$ is the convective flux of bound motors coming from a 
non-overlapping region, with $L$ being the arclength of a MT from its pole to
the overlapping region (Fig.~\ref{sketch}a). The value of $J(L,t)$ is 
determined by the solution of Eq.~\ref{motordyn}. When the motor
processivity length, $l_p\equiv 
V_0/k_u^0$, is smaller than the characteristic spindle length
($l_p\ll L$~\cite{values}), the flux $J(L,t)$ is determined by a constant bound motor density,
$\rho=k_b^{3D}\rho_{3D}/k_u^0$, fixed by the exchange of motors with the
bulk. In this case the dynamics of $n_c$ and $n_b$  are 
decoupled from $\rho$ close to the pole. 

The existence of antiparallel MT arrays under an external load $F$ is
determined by the balance between motor attachment and detachment fluxes, as
given by the steady state solutions, $\{n_c^f,n_b^f\}$, of
Eqs.~\ref{motordyn},~\ref{eq1}, which read
\begin{eqnarray}
\label{fixedpoints}
&&\delta  \equiv \frac{\rho_{3D} k_b^{3D} l_p}{\tilde{F}\alpha
k_u^0} \gamma = \frac{\exp \left( 1/\tilde{n}_c^f \right) (\tilde{n}_c^f)^2}{2 \left[ 1-\exp \left(
      -\tilde{L} \tilde{n}_c^f \right) \right]+\tilde{\ell} \tilde{n}_c^f} \;, \\
&&\tilde{n}_b^f = \frac{\exp \left( 1/\tilde{n}_c^f \right)
  \tilde{n}_c^f}{\gamma} \;. \nonumber
\end{eqnarray}
The quantities $\tilde{n}_c \equiv n_c / \tilde{F}$ and $\tilde{n}_b \equiv n_b
/ \tilde{F}$ are the normalized numbers of crosslinking
and bound motors respectively. The ratio, $\tilde{F}\equiv Fb/K_BT$,
between the force $F$ and the characteristic detachment force $K_BT/b$ sets
the natural scale of motors. The relevant dimensionless lengths are 
$\tilde{L}\equiv \alpha L/l_p$ and $\tilde{\ell} \equiv \alpha \ell/l_p$, and the
parameter $\alpha \equiv  f_s b/K_BT$ quantifies the sensitivity of
motor detachment to force. The asymmetry in motor attachment/detachment events at
vanishing load is characterized by $\gamma\equiv k_b/k_u^0$.

There always exists a critical value, $\delta_m$, below
which there are no solutions of Eq.~\ref{fixedpoints}. This situation
corresponds to an attachment flux of crosslinking motors that can not balance
their detachment flux, leading to the loss of all crosslinking motors and
inducing the spindle collapse. Associated to the critical point $\delta_m$, there 
is a minimum number of crosslinking motors, $\tilde{n}_c^{m}$, necessary to sustain a spindle,
whose value is given implicitly by
\begin{eqnarray}
&& 1+\left[ \tilde{n}_c^{m}\left(2+ \tilde{L}\tilde{n}_c^{m}\right) -1 \right]
 \exp \left( - \tilde{L}\tilde{n}_c^{m} \right) = \nonumber \\
&& \tilde{n}_c^{m}
  \left[ 2+\frac{\tilde{\ell}}{2} \left( \tilde{n}_c^{m}-1\right)\right] \;.
\label{minimmot}
\end{eqnarray}
The actual value of $\delta_m$ is obtained by substituting $\tilde{n}_c^{f}$ in 
Eq.~\ref{fixedpoints} by the solution of Eq.~\ref{minimmot}.

\begin{figure}
{\centering\resizebox*{8.5cm}{!}
  {\includegraphics{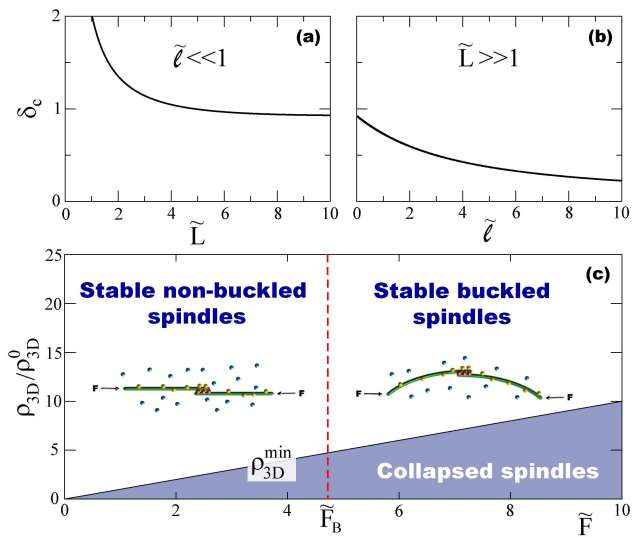}}}
\caption{\label{dynreg2} (a-b) Dependence
  of the critical value $\delta_c$ on $\tilde{L}$ and $\tilde{\ell}$ in
  the limiting cases where (a) $\tilde{\ell}\ll 1$  and (b) $\tilde{L}\gg 1$
  respectively. (c) Possible spindle structures as the bulk
  motor density $\rho_{3D}$ and the force $F$ are varied ($\rho_{3D}^0\equiv
  \delta_c \alpha (k_u^0)^3 / k_b^{3D} V_0 k_b$ sets the density units). Above
  $\rho_{3D}^{min}(F)$, buckled (straight) stable spindles exists for
  $F>F_B$ ($F<F_B$). Below $\rho_{3D}^{min}(F)$ no stable spindles exist.}
\end{figure}

In order to determine the stability of the structures, we perform a
linear stability analysis of the solutions of Eq.~\ref{fixedpoints}. For
$L\gg l_p$ it can be shown that the fluctuations in $\rho(s)$ are negligible and
the spindle stability depends only on the dynamics of $n_c$ and
$n_b$~\cite{fluctuations}. Stable spindles exist above a critical value
$\delta_c$. The precise expression for this 
critical point $\delta_c$ depends on the ratio between
motor attachment/detachment rates at vanishing load. For
$\gamma >\gamma_c \equiv  \exp \left(  1/\tilde{n}_c^{m}\right) \left(
  1-\tilde{n}_c^{m}\right)/\tilde{n}_c^{m}-1$, the transition from an unstable
array to a stable spindle corresponds to a saddle-node bifurcation at
$\delta_c=\delta_m$. On the other hand, if $\gamma <\gamma_c$, this transition
corresponds to a global bifurcation (saddle-connection~\cite{Holmes}) at a value $\delta_c >
\delta_m$.  Regardless the value of $\gamma$, the same  qualitative scenario is
observed as $\delta$ is varied and we restrict the following discussion to the
regime $\gamma > \gamma_c$ without loss of generality. The threshold value,
$\delta_c$, is fixed only by $\tilde{L}$ and $\tilde{\ell}$. Typically
$L\gg l_p$ and $\delta_c$ becomes independent of $L$ (Fig.~\ref{dynreg2}a),
while $\ell$ modifies $\delta_c$ slightly (Fig.~\ref{dynreg2}b). In this
limit, the spindle morphology and its stability are decoupled. As the motor  
properties and the lengths $L$ and $\ell$ are difficult to modify
experimentally, $\delta$ appears as the natural control
parameter for the spindle stability, since it depends both on the applied force,
$F$, and the bulk motor density, $\rho_{3D}$.

The existence of a critical value $\delta_c$  implies that for a
MT array under the action of a load, $F$, there exists a minimal
motor bulk concentration, $\rho_{3D}^{min}$, below which no stable spindles are
found. Using the definition of $\delta$ (Eq.~\ref{fixedpoints}), this minimal
motor density reads 
\begin{equation}
\rho_{3D}^{min} = \frac{k_u^0}{k_b^{3D}l_p}\frac{\alpha \delta_c}{\gamma}\tilde{F} \;.
\label{threshold}
\end{equation}
When $\rho_{3D}>\rho_{3D}^{min}$ the stable spindle may be either
straight or buckled depending on the value of the compressive force, $F$. For
forces below (above) the buckling force $F_B = (\pi/2)^2 
B/L^2$ of the structure ($B$ being its bending rigidity), the stable spindle
is straight (buckled). In Fig.~\ref{dynreg2}c we plot the structures that can
be found as the bulk density of motors and the force applied on the spindle
are varied. Indeed, recent experimental observations have shown that the progressive
inhibition of Eg5 motors leads to the collapse of the spindle at a finite bulk
motor density~\cite{Mitchison_eg5in}. Moreover, the total inhibition of
homolog motors (Klp61F) has been shown to prevent bipolar spindle formation
\emph{in vivo}~\cite{Vale}, in accordance with our predictions. 

\begin{figure}
{\centering\resizebox*{8.5cm}{!}
  {\includegraphics{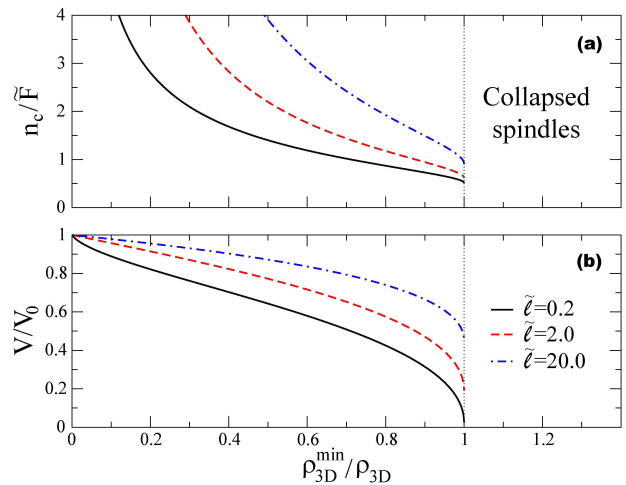}}}
\caption{\label{analitical} Steady state analytical solutions for (a) the number of
  crosslinking motors and (b) the velocity of the crosslinking motors (sliding
  velocity), as a function of the bulk density of motors. The different curves
  represent different values of $\tilde{\ell}$ in the limiting regime
  $\tilde{L}\gg 1$.}
\end{figure}

Above the threshold density $\rho_{3D}^{min}$,
there is a finite amount of crosslinking motors $n_c$ collectively holding the
spindle. In Fig.~\ref{analitical}a we plot the stable solutions
of Eq.~\ref{fixedpoints} as a function of the bulk
concentration of motors. Increasing values of $\rho_{3D}$ and $\ell$ leads to
larger motor attachment fluxes, that result in a larger amount of crosslinking
motors. For a living cell in native conditions, the 
MTs in the mitotic spindle are typically buckled~\cite{Salmon_Mitch}. Therefore, the force applied on
the MTs is of order $F_B$ which, for single
MTs ($5 \mu m$ in length) is about
$1pN$. Using this value for the force $F$, the number of crosslinking motors
leading to a stable antiparallel array turns out to be very small ($\simeq
2$). In this case fluctuations would dominate and, although stable arrays
could be transiently formed, their 
lifetime would be  too short (on the time scale of motor detachment). Since
the buckling force of a MT bundle can be at least one order of magnitude
larger~\cite{bundles}, stable interpolar MT bundles require tenths of crosslinking
motors and provide robust spindles with lifetimes over the time 
scale of the division process. Interpolar MT bundles are indeed observed in
several organisms during cell division~\cite{McIntosh01,McIntosh02,Sharp02}.  

The speed of the MT flux toward the poles is determined by the MT sliding
velocity, $\hat{V}$, given by the velocity of the crosslinking motors. In Fig.~\ref{analitical}b we
represent this sliding velocity as a function of $\rho_{3D}$,
for different values of $\tilde{\ell}$. It decreases from its maximal value $V_0$ as
the bulk motor density is decreased, and it is typically finite for the
minimal density $\rho_{3D}^{min}$ at which the spindle collapses, as observed experimentally~\cite{Mitchison_eg5in}.
At high motor concentrations $\rho_{3D}\gg \rho_{3D}^{min}$, the crosslinking
motors move nearly at their maximal velocity $V_0$ and the MTs move poleward
at this velocity consequently. As the 
motors in the non-overlapping regions move at velocity $V_0$ with respect to
the MTs in the spindle (Fig.~\ref{sketch}b), they appear static in the laboratory reference
frame. This apparent motor stillness has indeed been observed
experimentally~\cite{Mitchison_static}. Our analysis predicts that   
decreasing the bulk motor density $\rho_{3D}$ would allow the
observation of motor movement in the spindle. This observation
would provide further insight on the understanding of the mitotic
spindle structure.


The present approach highlights the importance of force-dependent
motor kinetics on the self-organization of MTs and motors.
In particular, we show that motor kinetics is a key factor in the
stability of spindle-like structures under applied load. Moreover, we have developed a
framework through which several observations in eukaryotic cell division, namely the spindle collapse, the MT
poleward flux, the static appearance of motors in the spindle, and
the existence of MT bundles, can be explained on a common physical ground. The
observation of motor motion upon a decrease of motor density, as predicted here, would provide strong
evidence for the self-organization of motors and MTs as the
underlying principle of mitotic spindle assembly.

\begin{acknowledgments}
We thank K. Kruse, J.-F. Joanny and J. Prost for stimulating discussions. We
thank the European Commission (HPRN-CT-2002-00312) and 
the Spanish M.E.C. (BQU2003-05042-C02-02 and FIS2005-01299) for financial support. I.P. also
thanks the {\sl Distinci\'o de la Generalitat de Catalunya} (Spain) for 
financial support.
\end{acknowledgments}


\begin{thebibliography}{99999}
\bibitem{Alberts} B. Alberts \emph{et al.}, \emph{The Molecular Biology of the Cell}
  (Garland, New York, 2002).
\bibitem{Nedelec01} F.J. Nedelec, T. Surrey, A.C. Maggs, S. Leibler, Nature \textbf{389}, 305 (1997).
\bibitem{Wittmann} T. Wittmann, A. Hyman, A. Desai, Nat. Cell Biol. \textbf{3}, E28 (2001).
\bibitem{Salmon_Mitch} T.J. Mitchison, E.D. Salmon, Nat. Cell Biol. \textbf{3}, E17 (2001).
\bibitem{Sharp} D.J. Sharp, G.C. Rogers, J.M. Scholey, Nature \textbf{407}, 41 (2000).
\bibitem{Kapoor} T.M. Kapoor, T.U. Mayer, M.L. Coughlin, T.J. Mitchison, J. Cell Biol. \textbf{150}, 975 (2000).
\bibitem{Mitchison_eg5in} D.T. Miyamoto \emph{et al.}, J. Cell Biol. \textbf{167}, 813 (2004).
\bibitem{Vale} G. Goshima, R.D. Vale, J. Cell Biol. \textbf{162}, 1003 (2003).
\bibitem{Sharp02} D.J. Sharp \emph{et al.}, J. Cell Biol. \textbf{144}, 125 (1999).
\bibitem{Schmidt} L.C. Kapitein \emph{et al.}, Nature \textbf{435}, 114 (2005).
\bibitem{Nicklas} R.B. Nicklas, J. Cell Biol. \textbf{97}, 542 (1983).
\bibitem{Mitchison_bipolar_flux} T.J. Mitchison \emph{et al.}, Mol. Biol. Cell \textbf{15}, 5603 (2004).
\bibitem{Frank} K. Kruse, F. J\"ulicher, Phys. Rev. Lett. \textbf{85}, 1778 (2000).
\bibitem{Jacques} K. Kruse, J.-F. Joanny, F. J\"ulicher, J. Prost,
  K. Sekimoto, Phys. Rev. Lett. \textbf{92}, 078101 (2004). 
\bibitem{assumption} This is reasonable for typical spindle lengths ($L_s\sim
  5\mu m$) as the motor bulk concentration equilibrates over
time scales, of order $\sim L_s^2/D\simeq 1 s$ ($D\sim 10\mu m^2 s^{-1}$ being the diffusion
constant of the motors in the bulk), shorter than the time scale of convective motor
movement along MTs, of order $\sim L_s/V_0\simeq 100 s$, for typical values of
the motor velocity ($V_0\simeq 33 nm s^{-1}$ for Eg5~\cite{Schmidt}).
\bibitem{Lipowsky} R. Lipowsky, S. Klumpp, T.M. Nieuwenhuizen,
  Phys. Rev. Lett. {\bf 87}, 108101 (2001). 
\bibitem{Parmeggiani} A. Parmeggiani, T. Franosch, E. Frey,
  Phys. Rev. Lett. {\bf 90}, 086601 (2003).
\bibitem{Block} M.J. Schnitzer, K. Visscher, S.M. Block, Nat. Cell
  Biol. \textbf{2}, 718 (2000).
\bibitem{VanKampen} N.G. Van Kampen, {\em Stochastic Processes in Physics and Chemistry}
(North Holland, Amsterdam, 2004).
\bibitem{Veigel} For non-processive myosin motors: C. Veigel, E.S. Molloy,
  S. Schmitz, J. Kendrick-Jones, Nat. Cell Biol. \textbf{5}, 980 (2003); for
  processive kinesin motors: Ref.~\cite{Block}. 
\bibitem{values}Typical values for $L$ are about
several microns long for most cell types, while Eg5 motors are not very
processive, $l_p<100nm$~\cite{Schmidt}.
\bibitem{fluctuations} For $L\ll l_p$  the stability scenarios remain
  qualitatively unchanged.
\bibitem{Holmes} J. Guckenheimer, P. Holmes, {\em Nonlinear
  Oscillations, Dynamical Systems, and Bifurcations of Vector Fields} (Applied
Mathematical Sciences, Vol. 42, Springer-Verlag, 1990). 
\bibitem{bundles} Assuming the bending rigidity of a MT bundle to scale
  lineraly with the number of MTs in the bundle, the buckling force of an interpolar
MT bundle may be of tenths of picoNewtons.
\bibitem{McIntosh01} D.N. Mastronarde, K.L. McDonald, R. Ding, J.R. McIntosh,
  J. Cell Biol. \textbf{123}, 1475 (1993).
\bibitem{McIntosh02} M. Winey \emph{et al.}, J. Cell Biol. \textbf{129}, 1601 (1995).
\bibitem{Mitchison_static} T.M. Kapoor, T.J. Mitchison, J. Cell
  Biol. \textbf{154}, 1125 (2001).
\end{thebibliography}
\end{document}